\documentclass[sigconf,authorversion,nonacm]{acmart}
\makeatletter
\def\@ACM@checkaffil{
    \if@ACM@instpresent\else
    \ClassWarningNoLine{\@classname}{No institution present for an affiliation}%
    \fi
    \if@ACM@citypresent\else
    \ClassWarningNoLine{\@classname}{No city present for an affiliation}%
    \fi
    \if@ACM@countrypresent\else
        \ClassWarningNoLine{\@classname}{No country present for an affiliation}%
    \fi
}
\makeatother
\usepackage{graphicx} 
\usepackage{subcaption}
\title{Distinguishing Scams and Fraud with Ensemble Learning}\titlenote{Authors are listed in alphabetical order.}
\author{Isha Chadalavada}
\affiliation{Northeastern University}
\author{Tianhui Huang}
\affiliation{Northeastern University}
\author{Jessica Staddon}
\affiliation{Northeastern University}
\thanks{Corresponding author: j.staddon@northeastern.edu}

\date{November 2024}

\begin{document}

\begin{abstract}
Users increasingly query LLM-enabled web chatbots for help with  scam defense. The Consumer Financial Protection Bureau's complaints database is a rich data source for evaluating LLM performance on user scam queries, but currently the corpus does not distinguish between scam and non-scam fraud. We developed an LLM ensemble approach to distinguishing scam and fraud CFPB complaints and describe initial findings regarding the strengths and weaknesses of LLMs in the scam defense context.
\end{abstract}
\maketitle

\section{Introduction}

A variety of large language model-based conversational assistants (LLMs) are available via the web and are regularly relied upon by users for distilling complex information and potentially reducing the work of information retrieval via search engines. An LLM use case of growing popularity is protection from scams and fraud; this use case is endorsed by security vendors \cite{zdnet_norton, biocatch, dataleon} and is the natural evolution of the user practice of seeking online advice for security-related tasks \cite{redmiles-2016-sources-n-selection}. The user-friendly nature and broad availability of web LLMs makes them a particularly compelling defense tool for socially isolated users, a group that is known to be vulnerable to scams (e.g., \cite{gimm2020financial,xing2020vulnerability}).

While the scam defense opportunities of LLMs are significant, the ability of the broadly accessible pretrained LLMs (e.g., ChatGPT, Google Gemini) to recognize scam markers has not been comprehensively assessed. In addition, LLMs are known to hallucinate information (e.g., \cite{GPT4SystemCard}) and come with privacy risks due to training data leaks \cite{carlini2021extracting} and a potential to encourage over-sharing of personal information by users \cite{zhang2024s}. To understand how to manage these risks in the scam context, a corpus of user scam narratives with which to evaluate LLM performance, is needed.

The Consumer Financial Protection Bureau (CFPB) complaints database \cite{cfpb}, is a promising source of scam narratives, however currently scam complaints are grouped with fraud complaints (through consumer selection of the ``fraud or scam'' issue or sub-issue when making a complaint). We have developed a  prompting technique \cite{sahoo2024systematic} for distinguishing CFPB complaints regarding a scam (i.e., a user has been \textit{tricked into taking} a financially self-harming action) from those concerning fraud (i.e., a financially harmful action is \textit{taken without the user's consent}). As part of the iterative development of this scam-identifying prompt, we have manually labeled a set of 300 CFPB complaints that consumers reported as ``fraud or scam'' complaints. We provide an analysis of this manually labeled set that suggests directions for improving the safety of  user interaction with LLMs in the scam context. 


In summary, we make two contributions in this short paper: 1. an ensemble prompting technique for distinguishing scam from fraud complaints in the CFPB database and 2. An analysis of LLM scam identification errors on a manually labeled set of 300 ``scam or fraud'' CFPB complaints, resulting in initial observations about the relationship between LLM performance and complaint length, redactions, and company names in complaint narratives. The second contribution may help developers fine-tune LLMs for better performance and help users more safely interact with LLMs. We plan to evaluate these observations on a larger set of narratives in future work.

\section{Background and Related Work}
\label{sec:background}

Consumer scams are a growing problem  in financial services (up $14\%$ in 2023 \cite{ftc_2023}). Many victims of scams (and other financial harms \cite{bruckner2023student,polat2023mortgage, cfpb_report_2023}) file complaints with the CFPB, a US financial sector consumer protection agency that  takes actions against financial services providers on behalf of consumers. With consumer consent, complaint narratives are made publicly accessible in redacted form \cite{cfpb_personal}. As of December 6, 2024 there are over 2.3 million narratives for which complainants have opted into inclusion in the public CFPB Complaints Database, almost 18,000 of which the complainant selected “fraud or scam” (the most granular scam-related annotation) as an issue or sub-issue. 

While scam and non-scam fraud narratives share  harms, the definitional distinction  is that in the case of scams, the user (or, complainant) takes self-harming actions \cite{modic2013scam}. In particular, while both types of complaints may involve transactions considered fraudulent, in the case of a scam a bad actor gains the user's trust allowing them to trick the user into authorizing the transaction themselves, whereas in non-scam fraud a bad actor authorizes the transaction (e.g., using stolen credentials). 

Human ability to recognize scams has long been studied (e.g., \cite{wash2020experts, cialdini2001science}), however, study of LLM ability to recognize scams is just emerging and to the best of our knowledge, all of the studies in this area (e.g., \cite{sehwag2024can,shen2024combating}) experiment with synthetic scam narratives. We are the first to provide a corpus of organic consumer scam narratives. The corpus is built by iteratively developing a high-precision ensemble LLM prompt for detecting scams, and so builds on previous research showing that with well-designed practices, LLMs can serve as good annotators (e.g., \cite{he-etal-2024-annollm, tan2024large}).

\subsection{Data and Ethics}
We plan to make our scam corpus publicly available, including the redacted narratives as they appear in the CFPB complaints database and the Gemini and GPT-4 responses to the prompts in Section~\ref{sec:final_prompt}. The CFPB redaction process \cite{cfpb_personal} is thorough, and to the best of our knowledge, none of the narratives include personally identifiable information. In addition, we have used the paid API service to access GPT-4, which ensures our queries are not used for future model development, as per OpenAI privacy policies.

\section{Prompt Development }
\label{sec:final_prompt}
We developed a prompt with high precision and recall through an iterative process of evaluating prompts against training data and improving the prompt according to observed error patterns. While we experimented with 3 LLMs, GPT-4\footnote{https://platform.openai.com/docs/models\#gpt-4}, Gemini\footnote{https://ai.google.dev/gemini-api/docs/models/gemini\#gemini-1.5-flash} and Llama\footnote{https://ollama.com/library/llama3.1}, our final prompt uses only Gemini and GPT-4 so we focus the discussion in this short paper on those 2 LLMs.

\paragraph{Training data.} To build a collection of complaints for training purposes, the authors independently labeled 150 CFPB complaints with the ``fraud or scam'' issue or subissue, relying on the definition of Section~\ref{sec:background} as the codebook (following standard qualitative coding practices, e.g., \cite{hruschka2004reliability}). Each complaint was labeled either ``scam'' or ``non-scam'', and each author added notes explaining their reasoning to facilitate later conversation. The authors met to discuss  their labels and were able to resolve all disagreements. The authors also agreed to take a ``customer is always right'' approach, meaning, that in the absence of evidence to the contrary, the complainaint's statements would be assumed correct and the complaint would be labeled accordingly. As an extreme example, if a complainant asserts they have been scammed and no further detail on what happened is provided, the complaint would be labeled ``scam''. Since we were able to establish high inter-related reliability through this process, each author then independently labeled 50 complaints, asking for second opinions as needed. As a result, our training data consisted of 300 CFPB complaints, $64\%$ of which are labeled scams. We denote this set of 300 labeled narratives by $L$. The narratives in $L$ range in character length from $38$ to 10,975 with a mean and median of 1,416.85 and 1,157.5, respectively. $93\%$ of the narratives in $L$ are somewhat redacted, and on average  $4.6\%$ of the narrative characters are redacted. 

\paragraph{Prompt Design and Iteration.} Each prompt we used relied heavily on the definition of a scam (Section~\ref{sec:background}) and required the LLMs to explain their labels since a large body of research shows performance can improve when LLM responses include descriptions of their ``reasoning'' (e.g. \cite{wei2022chain}). In addition, several prompts included example complaints and labels since examples have been associated with improved performance in other contexts (e.g., \cite{tan2024large}). Finally, we experimented with breaking prompts into multiple, simpler prompts (e.g., asking the LLM to just evaluate a single aspect of the definition and combining the LLM responses to decide on scam labels), and found this improved the performance of GPT-4.

We evaluated each prompt with Gemini and GPT-4 and calculated precision and recall. We also manually reviewed the errors and used the identified patterns to generate new prompts for evaluation. 

\paragraph{Final Prompt and Performance.} We achieved the best performance with an ensemble model using different prompts for Gemini and GPT-4 and requiring that the LLMs predict scam for each prompt. The prompts are:
\begin{enumerate}
\item Prompt A [used with Gemini]\footnote{Part of this prompt is an excerpt from https://en.wikipedia.org/wiki/Scam}: {\textit{``A scam is an attempt to defraud a person or group after first gaining their trust. Confidence tricks exploit victims using a combination of the victim's credulity, naïveté, compassion, vanity, confidence, irresponsibility, and greed. Researchers have defined confidence tricks as "a distinctive species of fraudulent conduct intending to further voluntary exchanges that are not mutually beneficial", as they "benefit con operators ('con men') at the expense of their victims (the 'marks')". Classify the following customer complaint as either "scam" or "not scam" (NO OTHER OPTIONS), then explain your decision. }}

\begin{itemize}
\item \textit{[Example 1] if the user inputs: ``I was asked to transfer money to an account after receiving a suspicious email. It claimed to be from my bank, but when I called my bank, they said they never sent such an email." The following answers should be generated: "Answer: scam Explanation: The customer describes a classic phishing attempt where a fraudulent email tries to convince them to send money, which was confirmed to be false by their bank."}
\item \textit{[Example 2] if the user inputs: "Eight unauthorized transactions totaling \$5800 were made from my account without my knowledge. I filed a claim with my bank, but they denied it twice, even after I provided proof of suspicious texts." The following answers should be generated: "Answer: not scam Explanation: The user describes unauthorized withdrawals from their bank account and attempts to recover the funds through their bank, which denied the claim. The user is likely a victim of financial fraud, not a scam perpetrator." }
\end{itemize}
\textit{Here is the input:[insert complaint]''}
\item Prompt B [used with GPT-4]: \textit{``Which of the following describes this complaint: \#1 Money was stolen from the complainant, \#2 The complainant was tricked into giving someone money, \#3 neither \#1 or \#2. Put your answer in format: Answer: [\#1/\#2/\#3] Explanation: [your explanation]} 
    
\textit{Here is the input: [insert complaint]''}
    
\item Prompt C [used with GPT-4]: \textit{``Which of the following describes this complaint: Potential Scam: The complainant gave money or personal information to a representative of a company of unknown reputation OR to someone impersonating a representative of a reputable company or organization; Not Scam: The complainant gave money or personal information to a representative of a reputable company or organization. Put your answer in format: Answer: [Potential Scam/Not Scam] Explanation: [your explanation]} 
    
\textit{Here is the input: [insert complaint]''} 
\end{enumerate}

The LLM responses to these prompts are combined via conjunction. That is, scam is predicted for a given complaint narrative if and only if: Gemini predicts scam in response to that narrative and prompt A  \textit{and} GPT-4 predicts \#2 for that narrative and prompt B \textit{and} GPT-4 predicts potential scam for that narrative and prompt C. We denote this ensemble model as $F$.  On the set $L$, $F$ achieves precision of $.95$ and recall of $.84$. 

When prompted with the $2569$ CFPB complaint narratives labeled with a ``fraud or scam'' issue or sub-issue from January 1, 2024 through November 13, 2024\footnote{At the time of writing, there are a little more than 2,700 narratives in the same time window, perhaps due to processing delays.}, 
$F$ identified 1,333 scam narratives ($.52$ of the data set). We manually evaluated a randomly selected 10\% sample ($n=133$), and measured a precision of $.97$. 

\begin{figure}
    \centering
    \includegraphics[width=.95\linewidth]{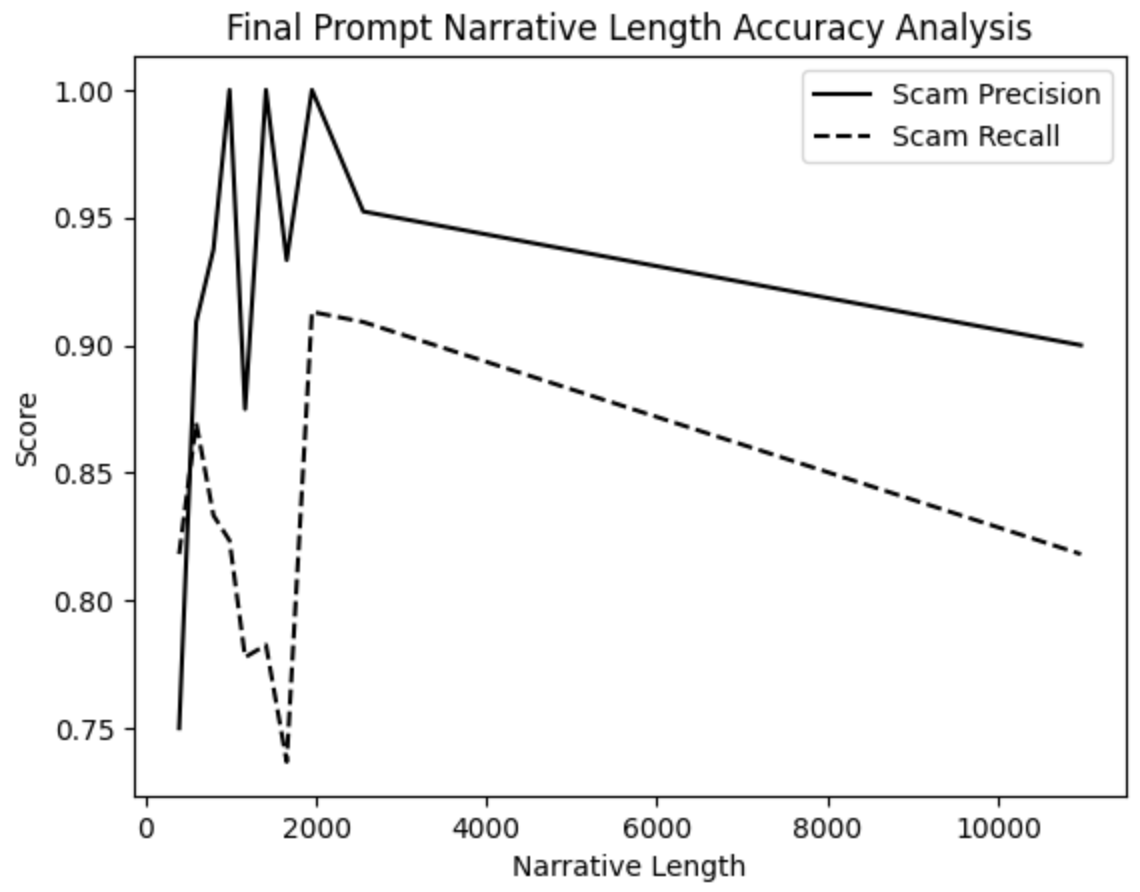}
    \caption{Precision and recall of the final model, $F$, as a function of complaint narrative length in characters.}
    \label{fig:performance_characters}
\end{figure}

\section{Error Patterns}
In this section we describe some of the patterns in the LLM errors that surfaced in responses to the labeled narratives, $L$.

\vspace{-5pt}
\paragraph{Reliance on Secondary Information} When complaint narratives are unclear and common scam markers are absent, we observed that both GPT-4 and Gemini can rely on secondary characteristics such as customer service or claim denials, to make decisions, rather than declining to decide. For example, CFPB Complaint \#4473515 devotes less than 14\% of its 950 characters to describe the financial harm that LLMs are determining to be a scam or non-scam fraud, and $6\%$ of the financial harm content is redacted: ``Someone took money from my Citibank account of My corporation and used it to pay XXXX XXXX i have no idea who these individuals are...'' The remaining $86\%$ of the complaint describes the complainant's experience trying to resolve the harm with Citibank.  While the narrative describes fraud (an unauthorized withdrawal) both GPT-4 and Gemini predict scam in response to prompt C and explain their decisions with secondary information. GPT says: ``The poor customer service experience combined with the unauthorized monetary transaction strongly suggests a potential scam.'' and Gemini responds: ``...the lack of communication from citibank, coupled with the unclear nature of the transaction and the involvement of a potentially fictitious individual or entity, points towards a potential scam.''
\vspace{-5pt}
\paragraph{Impact of Reputation} Scams are committed by entities of unknown or poor reputation and begin by building trust, thus persuading targets to act in the scammer's interest (e.g., \cite{shaari2019online}). Hence, a poor or unknown reputation is an important indicator of a potential scammer. We did not find evidence that LLMs consider reputation when identifying scams. Rather, there is some evidence of over-reliance on official business names as indicators of positive reputation. For example, the credit repair company, Lexington Law, has been fined by the CFPB for illegal fees and deceptive advertising and there is ample web evidence of its poor reputation (e.g., \cite{lexington}). Lexington Law is the subject of 4 complaints in the 300 labeled narratives, but only half of those are predicted scams with our final prompt, and some LLM explanations attribute a positive reputation to Lexington Law. In particular,  CFPB complaint \#4189755 begins ``Over the past year I have paid Lexington law firm and infinite finance to repair my credit. The negative items on my credit are old, settled, went through a chapter XXXX in 2012 or are simply not mine. I was told by both companies that it wouldn't be hard at all to get them removed...'' and both LLMs predicted non-scam; Gemini responded (via prompt A): ``The user's complaint describes a common experience with credit repair companies...'' and GPT-4 responded (via prompt C): ``The complainant mentions dealing with Lexington Law Firm and Infinite Finance, which are known and reputable companies...''. 

Particularly with GPT-4 we observed sensitivity to the language used to characterize the company in the narrative. For example, in CFPB narrative \#4308351 the complainant expresses some skepticism about Lexington Law (``My mother and I retained the Lexington Law Group, advertised and described as a credit repair agency.'') and GPT-4 with prompt C identified it as a potential scam and responded: ``Although the company is advertised as a credit repair agency, the nature of these issues might suggest that it could be a potential scam.''. 

Overall, we found better performance for narratives \textit{without} company names (recall of $.852$, precision of $.935$, with the ensemble model, $F$) than for narratives that include a company name (recall of $.733$, precision of $.846$ with the ensemble model, $F$). 
\vspace{-5pt}
\paragraph{Performance and Narrative Length} While precision initially improves with narrative length, for both GPT-4 and Gemini we found that model performance declines once narratives exceed approximately $3,000$ characters (Figure~\ref{fig:performance_characters}). Longer narratives often included information that is not directly relevant to determining whether the harm experienced was a scam or fraud, such as descriptions of the challenges the complainant encountered when trying to resolve the harm or cut and pastes of email correspondence. As mentioned earlier, this secondary information often surfaced in model explanations (incorrectly) justifying scam/fraud predictions. 

\begin{figure*}
    \centering
    \subfloat{{\includegraphics[width=4.9cm]{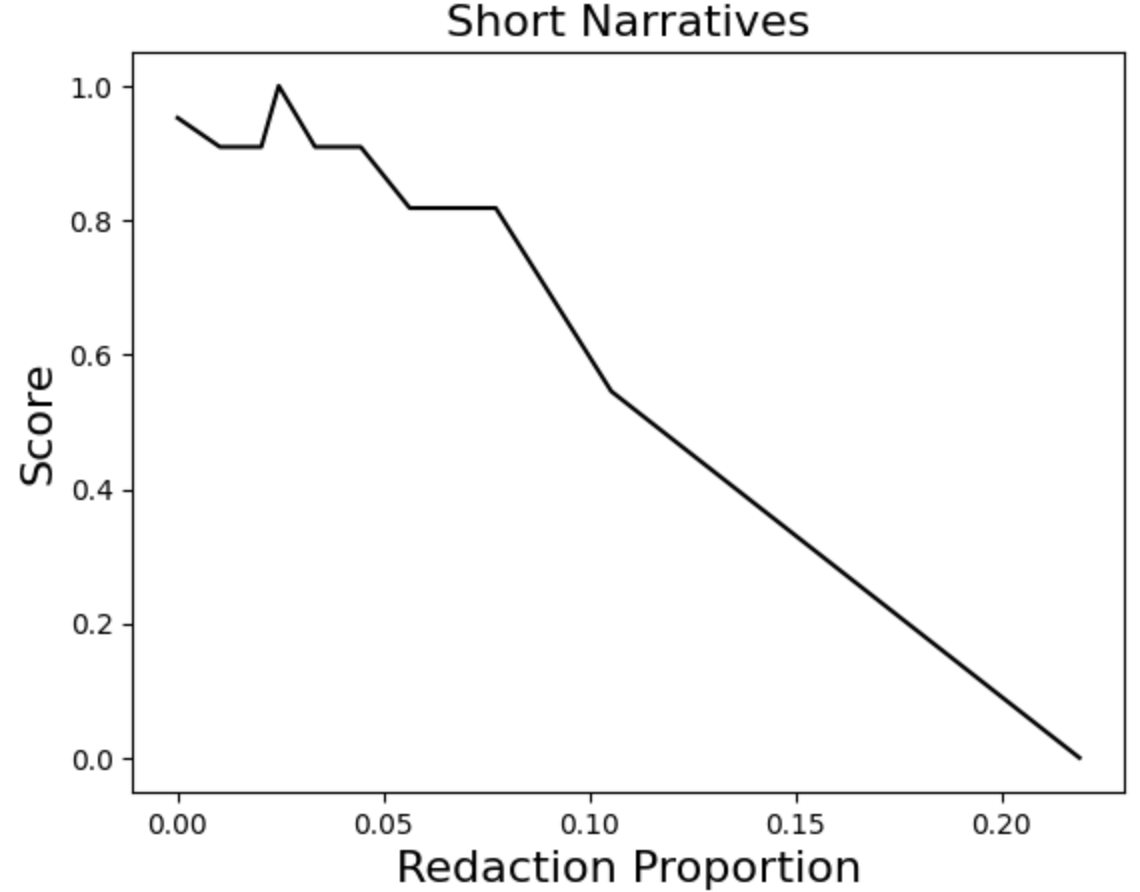}}}
    \qquad
    \subfloat{{\includegraphics[width=4.9cm]{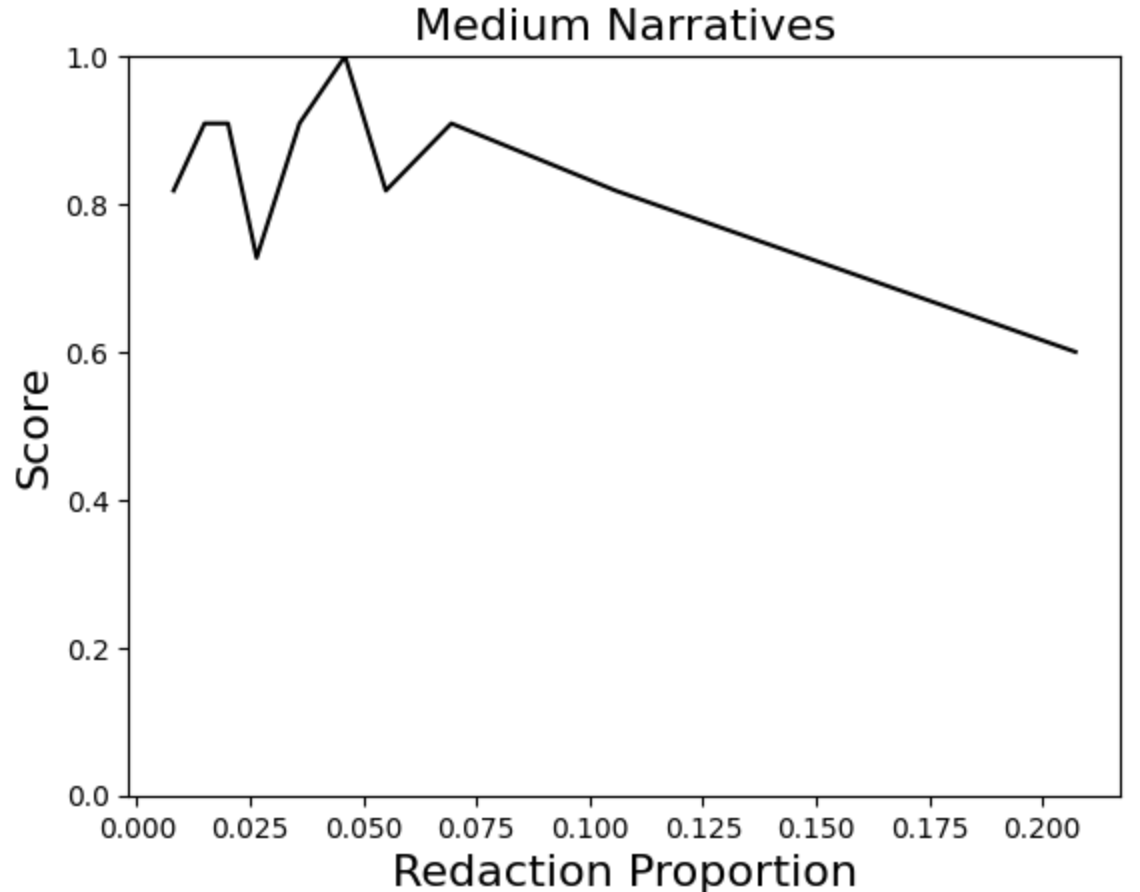}}}
    \qquad
    \subfloat{{\includegraphics[width=4.9cm]{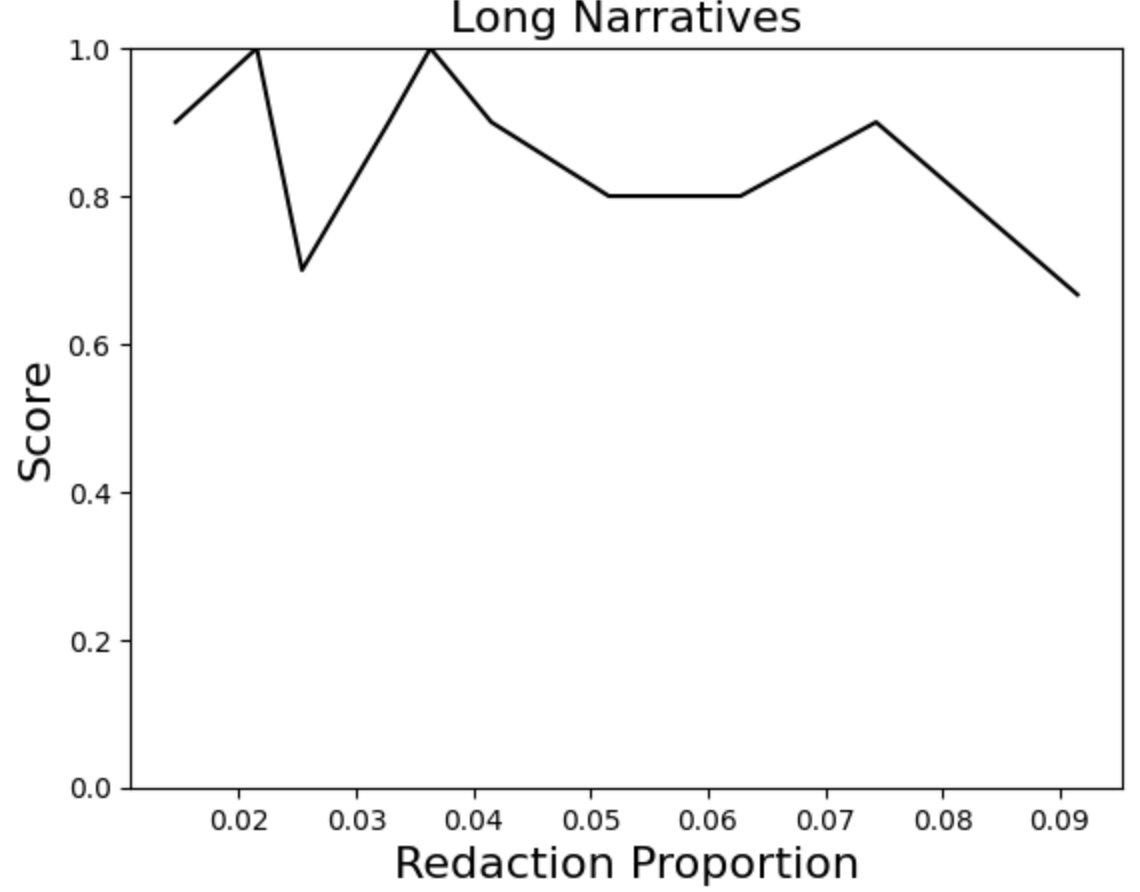}}}
    \caption{Each figure shows the accuracy of the final model, $F$, as a function of the fraction of redaction for narratives grouped by length. The narratives represented in the left most figure are the 99 messages in $L$ of at most 875 characters  (``short''); the middle figure represents the 100 narratives in $L$ of length 875 - 1,602 characters (``medium''); the right most figure represents the 101 narratives with at lest 1,602 characters (``long''). Note that performance is more robust to redaction with longer narratives.}
    \label{fig:performance_redaction}
\end{figure*}

\vspace{-5pt}
\paragraph{Performance on Redacted Narratives} Redaction removes information, and as is to be expected, negatively impacts LLM scam prediction performance. However, longer narratives can tolerate a higher percentage of redaction as shown in Figure~\ref{fig:performance_redaction}. We hypothesize that this is partly driven by the fact that LLMs  tend to predict scam when narratives assert a scam has occurred and the fraction of a complaint that is needed for such an assertion decreases as complaints grow. For example, CFPB complaint $\#4294462$ has more than $10\%$ redacted characters, but also includes an explicit scam declaration (``I soon realized that I was scammed and asked BB \& T to return the money....'') and is predicted a scam by $F$.

\section{Discussion and Open Problems}
We have developed an ensemble approach that uses Gemini and GPT-4 to distinguish scam from non-scam fraud CFPB complaints. We emphasize that the ensemble model is designed for the specific task of distinguishing scam from fraud and is not a general purpose scam complaint identifier. Indeed, we evaluated the prompt on a random subset of CFPB complaints that are not labeled with the ``fraud or scam'' issue or sub-issue  and found lower precision and recall ($F$ had a precision of $.25$ and recall of $.33$). 

Our evaluation surfaces several issues that need to be tested on a larger corpus and that we will explore in future work. We found evidence that LLMs take  secondary characteristics, for example, claim denials or the very existence of a scam/fraud complaint itself, as supporting evidence of scam. This has the potential to perpetuate biased decision making and may be due to the fact that, while consumer fraud protection is generally required, financial services institutions are less bound to compensate scam victims. It is important to understand the conditions under which LLMs rely on secondary characteristics to help consumers better express their concerns and help developers fine-tune models to improve performance.

The associations between complaint length and redaction proportion also have important implications for how consumers can use LLMs more effectively. In particular, our preliminary findings suggest that when using LLMs for scam identification, users should focus their narratives on their interactions with the potential scammer, and consider simulating or generalizing, rather than redacting, personal information in their narratives.

\bibliographystyle{plain}
\bibliography{pdasp_refs,references,usable-snp-related,llms-evals}
\end{document}